\documentstyle[preprint,aps,prb,epsf]{revtex}

\begin{document}

\draft

\title{
Spin-fluctuation exchange study of superconductivity in two- and
three-dimensional single-band Hubbard models
}

\author{
Ryotaro Arita, Kazuhiko Kuroki and Hideo Aoki
}
\address{Department of Physics, University of Tokyo, Hongo, Tokyo
113-0033, Japan}

\date{\today}

\maketitle

\begin{abstract}
In order to identify the most favorable situation for superconductivity 
in the repulsive single-band Hubbard model,
we have studied instabilities 
for $d$-wave pairing mediated by antiferromagnetic spin fluctuations 
and $p$-pairing mediated by ferromagnetic fluctuations 
with the fluctuation exchange approximation 
in both two dimensions and three dimensions.  
By systematically varying the band filling and band 
structure we have shown that 
(i) $d$-pairing is stronger in two dimensions than in three dimensions, and 
(ii) $p$-pairing is much weaker than the $d$-pairing.  
\end{abstract}

\medskip

\narrowtext
The discovery of the high-temperature superconductivity in
copper oxides by Bednorz and M{\" u}ller\cite{Bednorz} 
has kicked off intensive studies for electron mechanisms of superconductivity. 
Specifically, it is becoming increasingly clear that 
superconductivity can arise from repulsive electron-electron 
interactions.  A persuasive scenario is that
the superconductivity comes from a pairing interaction mediated by 
antiferromagnetic(AF) spin fluctuations.
A phenomenological calculation\cite{Moriya1,Moriya2,Moriya3,Pines1} 
along this line has 
succeeded in reproducing anisotropic $d$-wave superconductivity
as well as anomalous normal-state properties.  
Analytic calculations on a microscopic level 
with the fluctuation exchange approximation (FLEX), 
developed by Bickers {\it et al.}\cite{FLEX}, 
has also been applied to 
the Hubbard model on the two-dimensional(2D) square lattice
\cite{Dahm,Deisz} to show the occurrence of the superconductivity.  
Numerically, a quantum Monte Carlo study has indicated 
the pairing instability\cite{Kuroki}.  

These results indicate that 
the superconductivity near the AF instability in 2D 
has a `low $T_C$' $\sim O(0.01t)$ ($t$: transfer integral), 
i.e., two orders of magnitude smaller 
than the original electronic energy, but still 
`high $T_C$' $\sim O(100$ K) for $t\sim O(1$ eV).  
Then the next fundamental questions, which we address in this paper, are:
(i) Is 2D system more favorable for spin-fluctuation
mediated superconductivity than in three dimensions(3D)?
(ii) Can other pairing, such as a triplet $p$-pairing in the presence of 
{\it ferromagnetic} spin fluctuations, become competitive?
We take the single-band, repulsive Hubbard model as a 
simplest possible model, and look into the pairing with 
the FLEX method both in 2D and 3D.  The FLEX method 
has an advantage that systems having large spin fluctuations can be handled.  

Let us touch a little more upon the background to the above two 
questions.   The possibility of triplet pairing mediated by 
ferromagnetic fluctuations has been investigated
for superfluid $^3{\rm He}$\cite{Leggett}, 
a heavy fermion system ${\rm UPt_3}$\cite{heavyFermion}, and
most recently, an oxide ${\rm Sr_2RuO_4}$\cite{Sr}. 
It was shown that ferromagnetic fluctuations favor triplet pairing 
first by Layzer and Fay\cite{Layzer1}
before the experimental observation of p-wave pairing in $^3{\rm He}$.
For the electron gas model, Fay and Layzer\cite{Layzer2} or later
Chubukov\cite{Chubukov} has 
extended the Kohn-Luttinger theorem\cite{KohnLutt}
to $p$-pairing for 2D and 3D electron gas in the dilute limit.
Takada\cite{Takada} discussed the possibility of $p$-wave superconductivity
in the dilute electron gas with the Kukkonen-Overhauser model\cite{KO}.
As for lattice systems, 2D Hubbard model with large enough 
next-nearest-neighbor hopping $(t')$ has been shown to exhibit 
$p$-pairing for small band fillings.\cite{ChubukovLu} 
Hlubina\cite{Hlubina99} reached a similar conclusion by evaluating 
the superconducting vertex in a perturbative way.\cite{Takahashi}
However, the energy scale of the $p$-pairing in the Hubbard model, 
i.e., $T_C$, has not been evaluated so far.

As for 3D systems, Scalapino {\it et al}\cite{Scalapino}
showed for the Hubbard model that paramagnon
exchange near a spin-density wave instability gives rise to
a strong singlet $d$-wave pairing interaction, 
but $T_C$ was not discussed there.
Nakamura {\it et al}\cite{Nakamura}
extended Moriya's spin fluctuation theory of superconductivity\cite{Moriya2} 
to 3D systems, 
and concluded that $T_C$ is similar between the 2D and 3D cases 
provided that common 
parameter values (scaled by the band width) are taken. 
However, the parameters there are phenomelogical ones, 
so we wish to see whether the result remains valid 
for microscopic models.

Here we shall show that 
(i) $d$-wave instability mediated by AF 
spin fluctuation in 2D square lattice
is much stronger than those in 3D, while 
(ii) $p$-wave instability
mediated by ferromagnetic spin fluctuations in 2D are much weaker than 
the $d$-instability.  
These results, which cannot be predicted a priori, 
suggest that for the Hubbard model 
the `best' situation for the pairing instability 
is the 2D case with dominant AF fluctuations.

We consider the single-band Hubbard model 
with the transfer energy $t_{ij}=t (=1$ hereafter) for nearest 
neighbors along with $t_{ij}=t'$ for second-nearest neighbors, 
which is included to incorporate the band structure dependence.
The FLEX starts from a set of skeleton diagrams 
for the Luttinger-Ward functional to generate
a ($k$-dependent) self energy 
based on the idea of Baym and Kadanoff\cite{Baym}. 
Hence the FLEX approximation is a self-consistent perturbation
approximation with respect to on-site interaction $U$.

To obtain $T_C$, we solve, with the power method\cite{FLEX},
the eigenvalue ({\'E}liashberg) equation, 
\begin{eqnarray}
\lambda\Sigma^{(2)}(k)&=&\frac{T}{N}
\sum_{k'}
\Sigma^{(2)}(k')|G(k')|^2 V^{(2)}(k-k'),
\label{eliash}
\end{eqnarray}
where 
\begin{eqnarray}
V^{(2)}(q)&=&\frac{1}{2}
\left[ \frac{U^2\chi_0(q)}{1+U\chi_0(q)} \right]
-\frac{3}{2}
\left[ \frac{U^2\chi_0(q)}{1-U\chi_0(q)} \right]
\label{pair1}
\end{eqnarray}
for spin singlet pairing and
\begin{eqnarray}
V^{(2)}(q)&=&\frac{1}{2}
\left[ \frac{U^2\chi_0(q)}{1+U\chi_0(q)} \right] 
+\frac{1}{2}
\left[ \frac{U^2\chi_0(q)}{1-U\chi_0(q)} \right]
\label{pair2}
\end{eqnarray}
for spin triplet pairing,
where $\chi_0(q)\equiv -T/N\sum_k G(k)G(k+q)$ 
is the irreducible susceptibility, 
$G(k)$ the dressed Green's function, and $\Sigma^{(2)}(k)$ 
the anomalous self energy.  
At $T=T_C$, the maximum eigenvalue $\lambda_{\rm Max}$ reaches unity.
We take $N=64^2$ sites with $n_c=2048$ Matsubara frequencies for 2D, 
or $N=32^3$ with $n_c=1024$ for 3D.

\begin{figure}
\epsfxsize=7cm 
\epsfbox{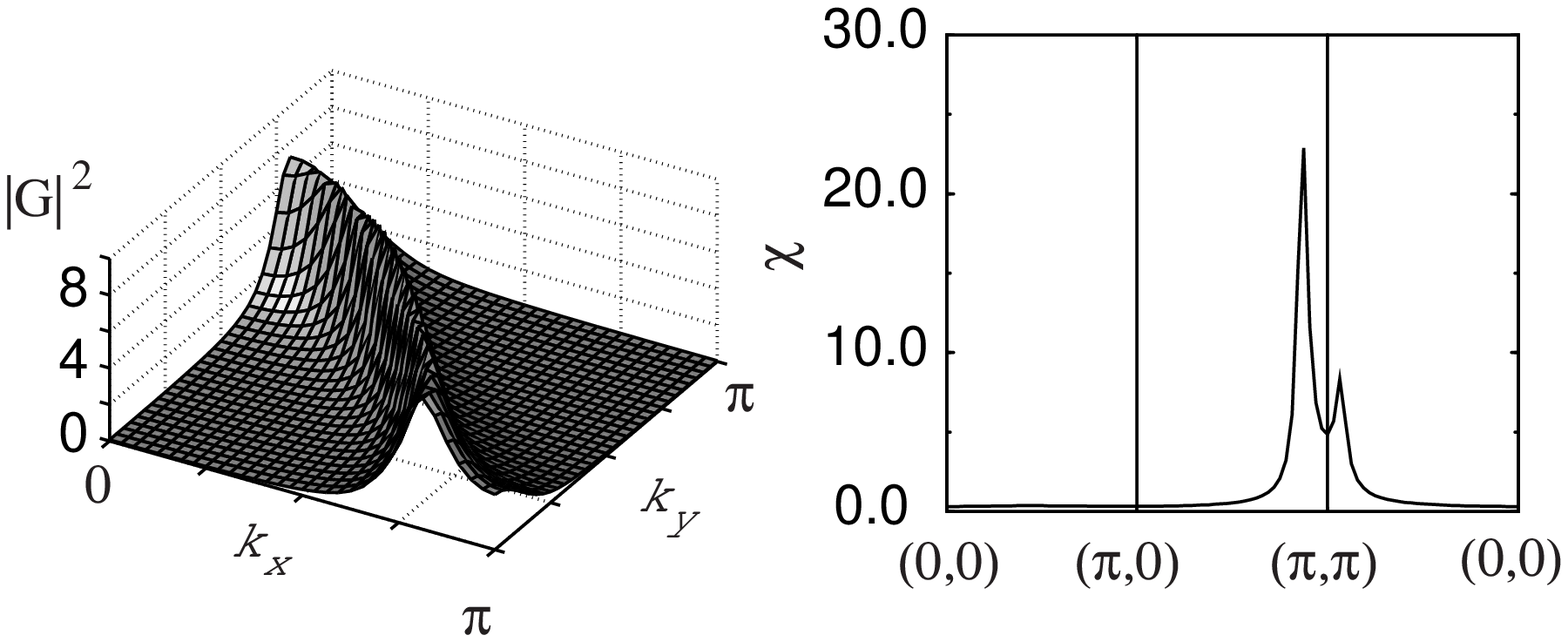}
\caption{
The squared 
absolute value of Green's function for the smallest Matsubara frequency, 
$i\omega_n=i\pi k_B T$ (left) and the RPA spin susceptibility (right) against 
the wave number for the 2D Hubbard model with $t'=0$, $n=0.85$ and
$U=4$.
}
\label{sl}
\end{figure}

Let us start with the 2D case having strong AF fluctuations.  
In Fig.\ref{sl}, we plot 
$\chi_{\rm RPA}(q) = \chi_0 /(1-U\chi_0)$ as a function of the momentum 
for the 2D Hubbard model with $t'=0$, $n=0.85$ (nearly half-filled) with 
$U=4$ and $T=0.03$. A dominant AF spin fluctuation is seen 
from $\chi_{\rm RPA}$ peaked near $(\pi,\pi)$.  

\begin{figure}
\epsfxsize=5cm 
\epsfbox{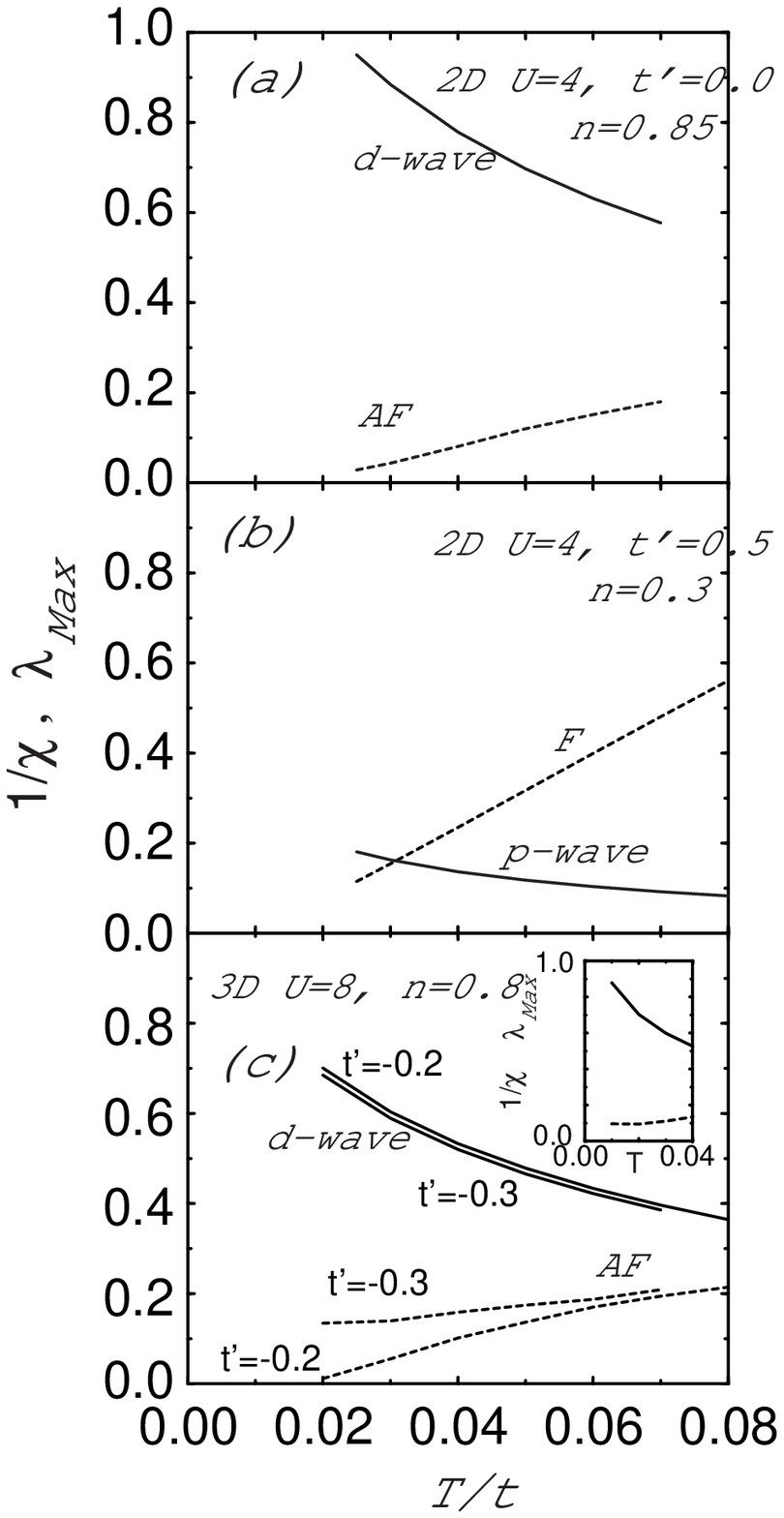}
\caption{
The maximum eigenvalue of the {\'E}liashberg equation 
(solid lines) 
and the reciprocal of the peak of $\chi_{\rm RPA}$ 
(either Ferro- or Antiferro-magnetic, dashed lines) 
against temperature for the Hubbard model in 
(a)2D with $t'=0$, $n=0.85$ and $U=4$,
(b)2D with $t'=0.5$, $n=0.3$ and $U=4$,
(c)3D with $t'=-0.2,-0.3$, $n=0.8$ and $U=8$.
The inset in (c) is the results for a larger 
number of Matsubara frequencies (=2048) for $t'=-0.3$.
}
\label{super}
\end{figure}

We can then solve the {\'E}liashberg equation (\ref{eliash}) to 
plot in Fig.\ref{super}(a) $\lambda_{\rm Max}$ as a function of 
temperature $T$ (normalized by $t$).  
The behavior of $|G(k, i\pi k_B T)|^2$ that appear in 
the {\'E}liashberg equation is indicated in Fig.\ref{sl}.  
How $\lambda_{\rm Max}$ is close to unity measures the pairing, 
and $\lambda_{\rm Max}$ tends to 
unity at $T\sim 0.02$, in accord with previous results\cite{Dahm,finitetc}.
We also plot the reciprocal of the peak value of 
$\chi_{\rm RPA}({\bf k},0)$, where $1/\chi \rightarrow 0$ indicates 
the magnetic ordering.  While we cannot compare $\lambda_{\rm Max}$ 
and $\chi_{\rm RPA}$ on an equal footing, since pairing fluctuations 
are neglected in the {\'E}liashberg equation while the susceptibility 
is treated beyond the mean field, we can discuss the behavior of 
$\lambda_{\rm Max}$ when the situation is varied. 

Keeping the above result in mind as a reference, we move on to
the case with ferromagnetic spin fluctuations, where 
triplet pairing is expected. This situation can be realized 
for relatively large $t'(\sim 0.5)$ 
and electron density away from half-filling in the 2D Hubbard model.
Physically, the van Hove singularity shifts toward the band bottom 
with $t'$, and the large density of states at the Fermi level 
for the dilute case favors the ferromagnetism.  
It has in fact been shown from quantum Monte Carlo study 
that the ground state is fully spin-polarized at 
$t'=0.47$, $n\sim 0.4$.\cite{Hlubina,Hlubina99}

\begin{figure}
\epsfxsize=7cm 
\epsfbox{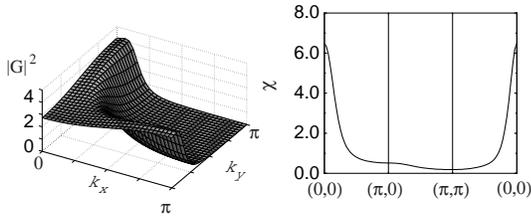}
\caption{
A similar plot as in Fig. \protect\ref{sl} 
for the 2D Hubbard model for a finite $t'=0.5$ with a smaller 
$n=0.3$ with $U=4$.
}
\label{tt'}
\end{figure}

We have calculated $\lambda_{\rm Max}$ 
for the density varied over $0.2\leq n \leq 0.6$ and 
$t'$ varied over $0.3 \leq t' \leq 0.6$ for $U=4, 6$ with $T=0.03$, 
and have found that $\lambda_{\rm Max}$ becomes largest for $n=0.3$, 
$t'=0.5$, so we concentrate on this parameter set hereafter.
If we look at in Fig. \ref{tt'} 
the momentum dependence of $|G(k, i\pi k_B T)|^2$
and $\chi_{\rm RPA}$ for this case with $U=4$, 
$\chi_{\rm RPA}$ is indeed peaked at $\Gamma$ (${\bf k}=(0,0)$).  
The question then is the behavior of 
$\lambda_{\rm Max}$ as a function of $T$, Fig.\ref{super}(b), which 
shows that $\lambda_{\rm Max}$ is much smaller than 
that in the AF case, Fig.\ref{super}(a).

A low $T_C$ for the ferromagnetic case 
contrasts with a naive expectation from the BCS picture, in which 
the Fermi level located around a peak in the density of states 
favors superconductivity.  
We may trace back two-fold reasons why this does not apply.  
First, if we look at the dominant ($\propto 1/[1-U\chi_0(q)]$) term of 
the pairing potential $V^{(2)}$ itself in eqs. (\ref{pair1}) and 
(\ref{pair2}), the triplet pairing interaction is only one-third of that
for singlet pairing.  
Second, the factor $|G|^2$ for the ferromagnetic case (Fig.\ref{tt'}) is 
smaller than that in the AF case (Fig.\ref{sl}), 
which implies that the self-energy correction is larger in the former.  
Larger self-energy correction (smaller $|G|^2$) leads to smaller 
eigenvalues of the {\'E}liashberg equation (\ref{eliash}). 
Even when we take a larger repulsion $U$ to increase the 
triplet pairing attraction (susceptibility), 
this makes the self-energy correction 
even stronger, resulting in only a small change in $\lambda$.

Let us now move on to the case of $d$-wave pairing 
in the 3D Hubbard model. 
In this case, we find that the $\Gamma_{3}^{+}$ representation of O$_{h}$
group\cite{Sigrist}
has the largest $\lambda_{\rm Max}$, so we look at this
pairing symmetry hereafter.
We have calculated $\lambda_{\rm Max}$ for the density 
varied over $0.75 \leq n \leq 0.9$ and 
$t'$ varied over $-0.5 \leq t' \leq +0.4$ for $U=4,6,8,10,12$ 
with $T=0.03$.
Among these parameter sets, we have found that $\lambda_{\rm Max}$ 
becomes largest for $n=0.8$, $t'=-0.2\sim -0.3$ and $U=8\sim 10$,
so hereafter we concentrate on this parameter set.

In Fig. \ref{super}(c), we again plot $\lambda_{\rm Max}$ along with 
the reciprocal of the peak value of $\chi_{\rm RPA}({\bf k},0)$ as a 
function of $T$ for $t'=-0.2,-0.3$ ,$U=8$ and $n=0.8$.
We can immediately see that the pairing tendency in 3D is much {\it weaker} 
than that in 2D.  Technically, for the sample size $N=32^3$ and the 
number of Matsubara frequencies $n_c=1024$ 
there are some finite-size effects 
for $T<0.02$.  As the inset for a larger $n_c=2048$ exemplifies, however, 
$\lambda_{\rm Max}$ tends to increase with $N$ and $n_c$, 
and we believe that a finite $T_C$ ($<0.01$) may be obtained at least
for $t'=-0.3$, $U=8$, $n=0.8$ in the limit of large $N$ and $n_c$, 
but this is still significantly smaller than in 2D.

Having confirmed this, the question now is: 
why is the $d$-superconductivity much stronger
in 2D than in 3D?  We can pinpoint the origin by looking at 
the various factors involved in the {\'E}liashberg equation. 
Namely we question the height of $V^{(2)}$ and $|G|^2$ 
along with the width of the region, both in the momentum sector 
and in the frequency sector, over which $V^{(2)}(k)$ 
contributes to the summation over $k\equiv ({\bf k},i\omega_n)$.

\begin{figure}
\epsfxsize=4cm 
\epsfbox{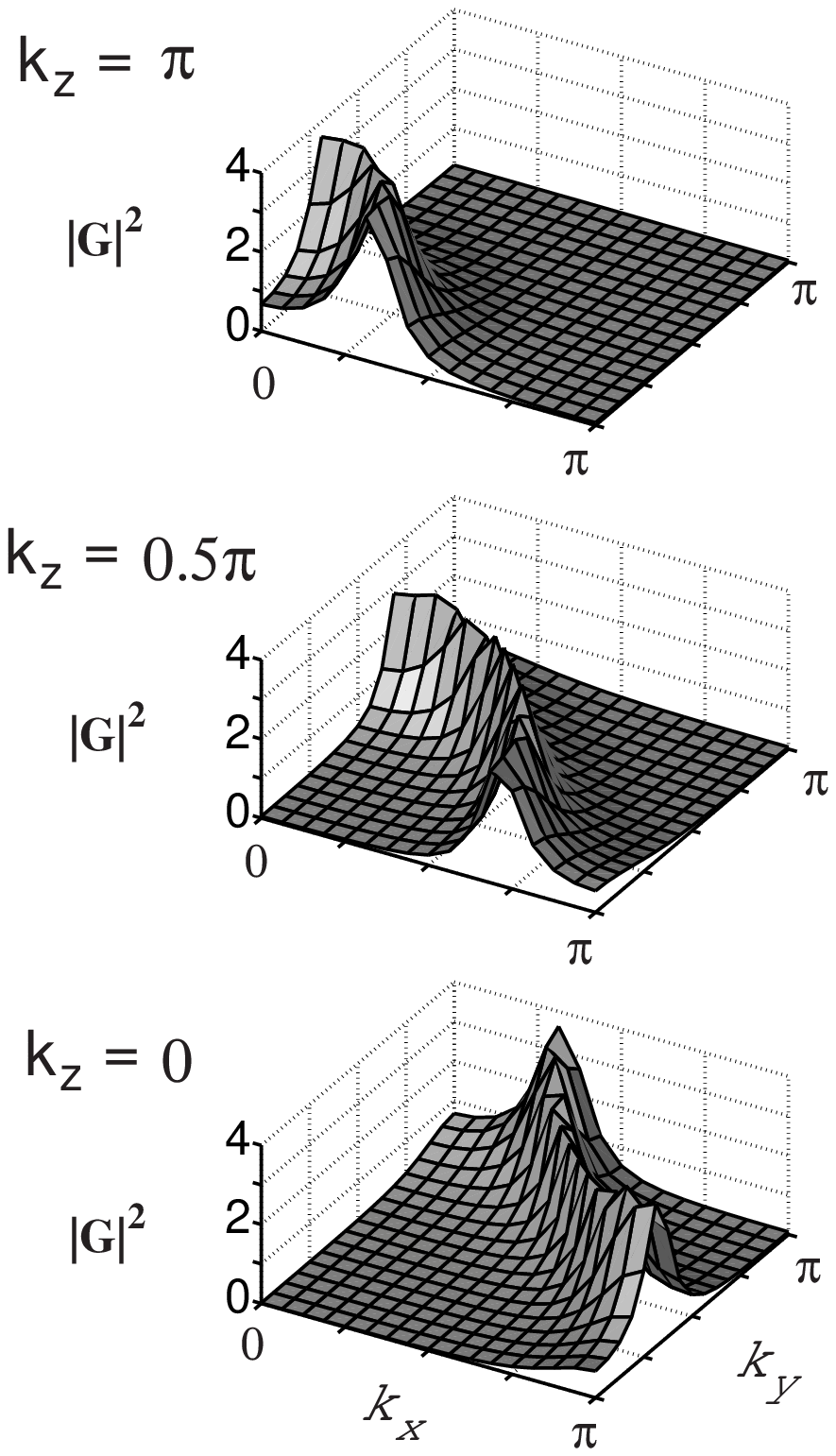}
\caption{
A plot for Green's function 
against $k_x$ and $k_y$ 
with $k_z=0, \pi/2, \pi$ 
for 3D Hubbard model with $t'=-0.2$, $n=0.8$ $U=8$, $T=0.03$. 
}
\label{sc1}
\end{figure}

We first plot $|G|^2$ for $k_z=0,\pi/2,\pi$ 
as a function of $k_x$ and $k_y$ 
in the 3D Hubbard model for $t'=-0.2$, $n=0.8$ with $U=8$ 
in Fig. \ref{sc1}.  We can see that the maximum of $|G|^2$ in 3D, 
if multiplied by $U^2$ arising in the {\'E}liashberg equation, 
is in fact larger than in 2D.  Were this factor the origin,
a larger $\lambda_{\rm Max}$ would result in 3D.

\begin{figure}
\epsfxsize=6cm 
\epsfbox{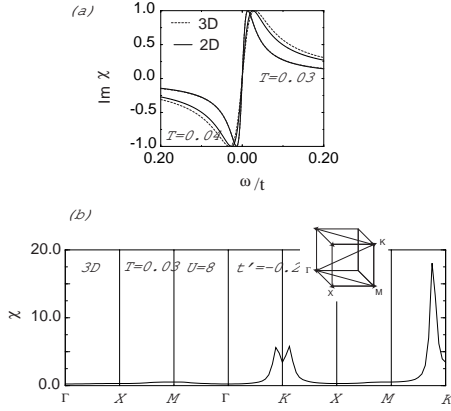}
\caption{
(a)${\rm Im}\chi_{\rm RPA}({\bf k_{Max}},\omega)$ 
(normalized by its maximum value)
as a function of $\omega/t$
for 3D Hubbard model with $t'=-0.2$, $n=0.8$, $U=8$ and $T=0.03,0.04$
(dashed line) and
for 2D Hubbard model with $t'=0$, $n=0.85$, $U=4$ and $T=0.03,0.04$ 
(solid line).
For $T=0.03$ 2D and 3D results almost overlap with each other.
(b)RPA spin susceptibility $\chi_{\rm RPA}({\bf k},0)$ 
as a function of the wave number for 3D Hubbard model with $t'=-0.2$, 
$n=0.8$, $T=0.03$ and $U=8$.
}
\label{sc2}
\end{figure}

We can then question how the peak in $\chi_{\rm RPA}$ spreads 
in the frequency axis. Fig. \ref{sc2}(a) 
displays ${\rm Im}\chi_{\rm RPA}(k_{\rm Max},\omega)$ 
(${\bf k}_{\rm Max}$: the momentum for which $\chi({\bf k},0)$ is maximum) 
as a function of $\omega$ (obtained by an analytic continuation 
with Pad{\'e} approximation\cite{Pade}).  
The figure compares the `best 3D' case ($t'=-0.2, n=0.8, U=8$) 
with a typical 2D case with $t'=0$, 
$n=0.85$ and $U=4$ having a similar magnitude of $\chi$.  
We can see that ${\rm Im}\chi(\omega)$, 
when this quantity is normalized by its maximum value 
while $\omega$ by $t$, 
exhibit surprisingly similar behaviors for 2D and 3D.  
So we can exclude the frequency width from the reason for the 
2D-3D difference. 
Note that if the frequency spread of the susceptibility 
scaled not with $t$ but with the {\it band width}, 
as Nakamura {\it et al}\cite{Nakamura}
have assumed, $\lambda_{\rm Max}$ would have become larger. 
So this is one reason why we stress that the present result 
that 2D is the best is by no means readily predictable.

If we turn to the momentum sector, 
Fig. \ref{sc2}(b) for $\chi_{\rm RPA}({\bf k},0)$ shows that 
the width, $a$, of the $\chi_{\rm RPA}({\bf k},0)$ peak in each 
momentum direction is similar to those in 2D (Fig.\ref{sl}).
Since the right-hand side of the {\'E}liashberg equation (\ref{eliash}) 
is normalized by $N \propto L^{D}$ 
with $L$ being the linear dimension of the system, 
$\lambda \propto (a/L)^D$ is smaller in 3D than that in 2D
when the main contribution of $V^{(2)}$ to $\lambda$ is confined 
around $(\pi,\pi)$ or $(\pi,\pi,\pi)$.
So we can conclude that this is the main reason why 2D differs from 3D.  

We have also obtained results (not shown here) in 3D for 
the body centered cubic lattice near half-filling 
(where strong AF fluctuations are expected), but the $d$-pairing is again weak. The $p$-pairing in the face centered cubic lattice with low band filling 
(where ferromagnetic fluctuations are expected) 
is found to be even weaker.  These results will be published elsewhere. 

To summarize, $d$-pairing in 2D is the best situation for the repulsion 
originated (i.e., spin fluctuation mediated) superconductivity
in the Hubbard model.  
In this sense, the layer-type cuprates do seem to hit upon the 
right situation.  
However, our conclusion has been obtained for the simplest possible 
single-band Hubbard model, while the
detailed behavior of $T_C$ may depend on the model. 
Indeed, if we turn to other 3D superconductors,
the heavy fermion system, in which the pairing 
is thought to be meditated by spin fluctuations,
the $T_C$, when normalized by the band width $W$, 
is known to be of the order of $0.001W$.   
Since the present result indicates that $T_C$, normalized by $W$, 
is $\sim 0.0001W$ at best in the 3D Hubbard model, 
we may envisage that the heavy fermion system is an instance 
in which larger frequency and/or momentum spreads in 
$\chi({\bf k},\omega)$ are utilized than in the Hubbard model.

After completion of this study, we came to know the work
by Monthoux and Lonzarich.\cite{MonLon}
Using a phenomenological approach, they conclude for 2D systems 
that the $d$-wave pairing is much stronger than $p$-wave pairing,
which is consistent with the present result.

We would like to thank K. Ueda and H. Kontani for illuminating 
discussions.
R.A. would like to thank S. Koikegami for discussions on the FLEX.
R.A. is supported by a JSPS Research Fellowship for
Young Scientists, while 
K.K. acknowledges a Grant-in-Aid for Scientific
Research from the Ministry of Education of Japan.
Numerical calculations were performed at the Supercomputer Center,
ISSP, University of Tokyo.

\begin{references}
\bibitem{Bednorz}J. G. Bednorz and K. A. M{\" u}ller, Z. Phys. B {\rm 64},
189 (1986).
\bibitem{Moriya1}T. Moriya, Y. Takahashi, and K. Ueda,
J. Phys. Soc. Jpn, {\bf 59}, 2905 (1990); 
Physica C {\bf 185-189}, 114 (1991).
\bibitem{Moriya2}T. Ueda, T. Moriya, and Y. Takahashi,
{\it Electronic Properties and Mechanisms of High-$T_C$ Superconductors} 
ed. T. Oguchi {\it et al.} (North Holland, Amsterdam, 1992), p. 145; 
J. Phys. Chem. Solids {\bf 53}, 1515 (1992).
\bibitem{Moriya3}T. Moriya and K. Ueda, J. Phys. Soc. Jpn. {\bf 63}, 
1871, (1994).
\bibitem{Pines1}P. Monthoux, A. V. Balatsky, and D. Pines,
Phys. Rev. B {\bf 46}, 14803 (1992); 
P. Monthoux and D. Pines,
Phys. Rev. B {\bf 47}, 6069 (1993); 
{\it ibid} {\bf 49}, 4261 (1994).
\bibitem{FLEX}N. E. Bickers, D. J. Scalapino, and S. R. White,
Phys. Rev. Lett. {\bf 62}, 961 (1989); 
N. E. Bickers and D. J. Scalapino, Ann. Phys. (N. Y.)
{\bf 193}, 206 (1989).
\bibitem{Dahm}T. Dahm and L. Tewordt, Phys. Rev. B {\bf 52}, 1297 (1995).
\bibitem{Deisz}J.J. Deisz, D. W. Hess, and J. W. Serene,
Phys. Rev. Lett. {\bf 76}, 1312 (1996).
\bibitem{Kuroki}K. Kuroki and H. Aoki, Phys. Rev. B {\bf 56} R14287(1997);
K. Kuroki and H. Aoki, J. Phys. Soc. Jpn {\bf 67}, 1533 (1998).
\bibitem{Leggett}A. J. Leggett, Rev. Mod. Phys. {\bf 47}, 331 (1975).
\bibitem{heavyFermion}H. Tou {\it et al.}, 
Phys. Rev. Lett. {\bf 77}, 1374 (1996);
{\it ibid}, {\bf 80}, 3129 (1998).
\bibitem{Sr}Y. Maeno {\it et al.}, 
Nature {\bf 372}, 532 (1994); 
T. M. Rice, M. Sigrist, J. Phys. Condens. Matter {\bf 7}, L643 (1995).
\bibitem{Chubukov}M. Yu. Kagan and A. V. Chubukov, Pis'ma Zh. Eksp.
Teor. Fiz. {\bf 47}, 525 (1988); 
A. V. Chubukov, Phys. Rev. B {\bf 48}, 1097 (1993).
\bibitem{Layzer1}A. Layzer and D. Fay, Int. J. Magnetism 1, 135 (1971);
Proc. Of the IIth. Int. Conf. On Low Temp. Physics (LTI 1), 
St. Andrews Press(1968), Vol.2, page 760.
\bibitem{Layzer2}D. Fay and A. Layzer, Phys. Rev. Lett. {\bf 20},
187(1968)
\bibitem{KohnLutt}W. Kohn and J.M. Luttinger, Phys. Rev. Lett. {\bf 15}, 
524 (1965).
\bibitem{Takada}Y. Takada, Phys. Rev. B {\bf 47}, 5202 (1993).
\bibitem{KO}C. A. Kukkonen and A. W. Overhauser, 
Phys. Rev. B {\bf 20} 550(1979).
\bibitem{ChubukovLu}A. V. Chubukov and J. P. Lu, Phys. Rev. B {\bf 46}, 
11163 (1992).
\bibitem{Hlubina99}R. Hlubina, Phys. Rev. B {\bf 59}, 9600 (1999).
\bibitem{Takahashi} 
H. Takahashi [J. Phys. Soc. Jpn, {\bf 68}, 194 (1999)] 
on the other hand concludes that $p$-wave channel is most attractive 
for dilute ($n\sim 0.1$) 2D Hubbard model on the $t'=0$ square lattice.
\bibitem{Scalapino}D. J. Scalapino, E. Loh, Jr., and 
J. E. Hirsh, Phys. Rev. B {\bf 34}, 8190 (1986).
\bibitem{Nakamura}S. Nakamura, T. Moriya and K. Ueda,
J. Phys. Soc. Jpn {\bf 65}, 4026 (1996).
\bibitem{Baym}G. Baym and L.P. Kadanoff, Phys. Rev. {\bf 124}, 
287 (1961); G. Baym, Phys. Rev. {\bf 127}, 1391 (1962).
\bibitem{finitetc} Here 
a finite $T_C$ in 2D systems is thought of as a measure of $T_C$ 
when the layers are stacked to Josephson-couple.  
\bibitem{Hlubina}R. Hlubina, S. Sorella, and F. Guinea, 
Phys. Rev. Lett. {\bf 78}, 1343 (1997).
\bibitem{Sigrist}M. Sigrist and K. Ueda, Rev. Mod. Phys., 63, 239 (1991).
\bibitem{Pade}H. J. Vidberg and J. W. Serene, J. Low. Temp. Phys.
{\bf 29}, 179 (1977).
\bibitem{MonLon}P. Monthoux and G. G. Lonzarich,
Phys. Rev. B {\bf 59}, 14598 (1999).
\end{references}
\end{document}